\documentstyle[preprint,aps,epsf]{revtex}% Use this for style revtex

\begin{document}
\def\beq{\begin{equation}}
\def\eeq{\end{equation}}
\def\bea{\begin{eqnarray}}
\def\eea{\end{eqnarray}}
\def\oupb{UPB\ }
\def\pb{PB\ }
\def\eps{\epsilon}
\newcommand{\ket}[1]{| #1 \rangle}
\newcommand{\bra}[1]{\langle #1 |}
\newcommand{\braket}[2]{\langle #1 | #2 \rangle}
\newcommand{\proj}[1]{| #1\rangle\!\langle #1 |}
\newcommand{\ba}{\begin{array}}
\newcommand{\ea}{\end{array}}
\newtheorem{theo}{Theorem}
\newtheorem{defi}{Definition}
\newtheorem{lem}{Lemma}
\newtheorem{exam}{Example}
\newtheorem{propo}{Proposition}

\author{Barbara M. Terhal}

\title{A Family of Indecomposable Positive Linear Maps based on 
Entangled Quantum States}

 \address{\vspace*{1.2ex}
            \hspace*{0.5ex}{Instituut voor Theoretische Fysica, Universiteit van 
Amsterdam, Valckenierstraat 65, 1018 XE Amsterdam, and Centrum voor Wiskunde en Informatica, Kruislaan 413, 1098 SJ Amsterdam, The Netherlands.}\\
Email: {\tt terhal@wins.uva.nl},{\tt terhal@watson.ibm.com}}

\date{\today}

\maketitle
\begin{abstract}
We introduce a new family of indecomposable positive linear maps based on entangled quantum states. Central to our construction is 
the notion of an unextendible product basis. The construction lets us create indecomposable positive linear maps in matrix 
algebras of arbitrary high dimension.
\end{abstract}

%
%\narrowtext
%\widetext

\section{Introduction}
\label{intro}

One of the central problems in the emergent field of 
quantum information theory \cite{qit} is the classification and 
characterization of the entanglement (to be defined in section \ref{fampos}) of quantum states. Entangled 
quantum states have been shown to be valuable resources in (quantum) 
communication and computation protocols. In this context it has 
been shown \cite{nec_horo} that there exists a strong connection between 
the classification of the entanglement of quantum states and the structure of 
positive linear maps. Very little is known about the structure of 
positive linear maps even on low dimensional matrix algebras, in particular 
the structure of indecomposable positive linear maps. We denote the $n \times n$ matrix algebra as $M_n({\bf C})$. The first example of an 
indecomposable positive linear map in $M_3({\bf C})$ was found by Choi 
\cite{biqua}. There have been only several other examples of indecomposable 
positive linear maps (see Ref. \cite{hosaka} for some recent literature); they seem to be hard to find and no general construction method is available.
In this paper we make use of the connection with quantum states
to develop a method to create indecomposable positive linear maps on 
matrix algebras $M_n({\bf C})$ for any $n > 2$. Central in this construction 
is the notion of an unextendible product basis, of which there exist examples in 
arbitrary high dimensions \cite{upb,upbbe2}. Unextendible product bases have turned out to be mathematically rich objects which can be understood with the use of graph theoretic and linear algebraic tools \cite{lovalon}.

In section \ref{fampos} we present the general construction. In 
section \ref{exampl} we present two examples and discuss various open 
problems.

\section{Unextendible Product Bases and Indecomposable Maps}
\label{fampos}

A $n$-dimensional complex Hilbert space is denoted as ${\cal H}_n$.
The set of all linear operators on a Hilbert space ${\cal H}_n$ will be denoted as $B({\cal H}_n)$. The subset of Hermitian positive 
semidefinite operators is denoted as $B({\cal H}_n)^{+}$. We will use the conventional bra and ket notation in 
quantum mechanics, i.e. a vector $\vec{\psi}$ in ${\cal H}_n$ is written as 
a ket,  
\beq
\ket{\psi} \in {\cal H}_n, 
\eeq
and the Hermitian conjugate of $\vec{\psi}$, $\vec{\psi}^*$, is denoted as a 
bra $\bra{\psi}$. The complex innerproduct between vectors $\ket{\psi}$ and 
$\ket{\phi}$ in ${\cal H}_n$ is denoted as
\beq
\langle \psi| \phi \rangle \equiv \vec{\psi}^{*} \vec{\phi}.
\eeq
The vectors $\ket{\psi} \in {\cal H}$ are usually normalized, $\langle \psi| \psi \rangle=1$. Elements of $B({\cal H}_n)^{+}$ are 
denoted as
\beq
\rho=\sum_i \lambda_i \ket{\psi_i} \bra{\psi_i},
\eeq
where $\ket{\psi_i}$ are the normalized eigenvectors of $\rho$ and $\lambda_i \geq 0$ 
are the eigenvalues. When $\rho$ has trace equal to one, $\rho$ is said to be a density matrix. The physical state of a quantum mechanical
system is given by its density matrix. If a density matrix $\rho$ has rank one, $\rho$ is 
called a pure state and can be written as 
\beq
\rho=\ket{\psi}\bra{\psi}.
\eeq

Let ${\cal S} \colon \; B({\cal H}_n) \rightarrow B({\cal H}_m)$ be a 
linear map. ${\cal S}$ is positive when ${\cal S} \colon \; B({\cal H}_n)^+ \rightarrow B({\cal H}_m)^+$. Let ${\rm id}_k$ be the identity map on $B({\cal H}_k)$. We define the map 
${\rm id}_k \otimes {\cal S}\colon\; B({\cal H}_k \otimes {\cal H}_n) 
\rightarrow B({\cal H}_k \otimes {\cal H}_m)$ for $k=1,2,\ldots$ by  
\beq
\left({\rm id}_k \otimes {\cal S}\right) \left(\sum_i \sigma_i \otimes \tau_i\right)=\sum_i \sigma_i \otimes 
{\cal S}(\tau_i),
\eeq
where $\sigma_i \in B({\cal H}_k)$ and $\tau_i \in B({\cal H}_n)$.
The map ${\cal S}$ is $k$-positive when ${\rm id}_k \otimes 
{\cal S}$ is positive. The map ${\cal S}$ is completely positive when ${\cal S}$ is $k$-positive for all $k=1,2,\ldots$.
Following Lindblad \cite{lindblad}, the set of physical operations on a density matrix $\rho \in B({\cal H}_n)^+$ is given by the set of completely positive 
trace-preserving maps ${\cal S} \colon \;B({\cal H}_n) \rightarrow B({\cal H}_m)$.
Similarly as $k$-positive, one can define a $k$-copositive map. Let 
$T\colon B({\cal H}_n) \rightarrow B({\cal H}_n)$ be defined as matrix 
transposition in a chosen basis for ${\cal H}_n$, i.e. 
\beq
(T(A))_{ij}=A_{ji},
\eeq
on a matrix $A \in B({\cal H}_n)$. The map ${\cal S}$ is $k$-copositive 
when ${\rm id}_k \otimes ({\cal S} \circ T)$ is positive. A positive 
linear map ${\cal S} \colon \; B({\cal H}_n) \rightarrow B({\cal H}_m)$ is 
decomposable if it can be written as 
\beq
{\cal S}={\cal S}_1+{\cal S}_2 \circ T,
\eeq
where ${\cal S}_1\colon \; B({\cal H}_n) \rightarrow B({\cal H}_m)$ and 
${\cal S}_2\colon \; B({\cal H}_n) \rightarrow B({\cal H}_m)$ are completely positive maps.  
It has been shown by Woronowicz \cite{woro} that all positive linear 
maps ${\cal S}\colon \; B({\cal H}_2) \rightarrow B({\cal H}_2)$ and 
${\cal S}\colon \; B({\cal H}_2) \rightarrow B({\cal H}_3)$ are decomposable.

\begin{defi}
Let $\rho$ be a density matrix on a finite dimensional Hilbert space 
${\cal H}_A \otimes {\cal H}_B$. A state $\ket{\psi}$ 
of the form $\ket{\psi^A} \otimes \ket{\psi^B}$ is a (pure) product state in 
${\cal H}_A \otimes {\cal H}_B$. The density matrix $\rho$ is entangled iff $\rho$ cannot be written as a nonnegative combination of pure product states, i.e. there does not exist an ensemble 
$\{p_i \geq 0,\ket{\psi_i^A \otimes \psi_i^B}\}$ 
such that
\beq
\rho=\sum_i p_i \,\ket{\psi_i^A}\bra{\psi_i^A} \otimes \ket{\psi_i^B}\bra{\psi_i^B}.
\eeq 
When $\rho$ is not entangled $\rho$ is called separable. 
\end{defi}

The problem of deciding whether a bipartite density matrix $\rho$ on 
${\cal H}_A \otimes {\cal H}_B$ is entangled can be quite hard. 
The following theorem by the Horodeckis \cite{nec_horo} formulates a necessary 
and sufficient condition for a density matrix $\rho$ to be entangled:

\begin{theo}[Horodecki]
A density matrix $\rho$ on ${\cal H}_A \otimes {\cal H}_B$ is entangled iff
there exists a positive linear map ${\cal S}\colon\;{\cal H}_B \rightarrow 
{\cal H}_A$ such that 
\beq
\left({\rm id}_{A} \otimes {\cal S}\right)(\rho) 
\eeq
is not positive semidefinite. Here ${\rm id}_A$ denotes the identity map on 
$B({\cal H}_A)$.
\label{horothm}
\end{theo}

{\em Remark} An equivalent statement as Theorem \ref{horothm} holds for
positive linear maps ${\cal S}\colon\;{\cal H}_A \rightarrow {\cal H}_B$ and 
the positivity of ${\cal S} \otimes {\rm id}_{B}$.

The consequences of Theorem \ref{horothm} and Woronowicz' result is that 
a bipartite density matrix $\rho$ on ${\cal H}_2 \otimes {\cal H}_2$ and 
${\cal H}_2 \otimes {\cal H}_3$ is entangled iff $\left({\rm id}_{A} \otimes {\cal S}_1+{\cal S}_2 \circ T\right)(\rho)$ is not positive semidefinite.
As ${\cal S}_1$ and ${\cal S}_2$ are completely positive maps this is 
equivalent to saying that $\left({\rm id}_{A} \otimes T\right)(\rho)$ is not 
positive semidefinite.

%which was conjectured by Peres \cite{peres} for any 
%bipartite state) 

The more complicated structure of the positive linear maps in higher 
dimensional matrix algebras, namely the existence of indecomposable
positive maps, is reflected by the existence of entangled density 
matrices $\rho$ on ${\cal H}_A \otimes {\cal H}_B$ for which 
$\left({\rm id}_{A} \otimes T\right)(\rho)$ {\em is} positive semidefinite.
%The entanglement of these density matrices is called bound.

The first example of such a density matrix on ${\cal H}_2 \otimes 
{\cal H}_4$ and ${\cal H}_3 \otimes {\cal H}_3$ was found by 
P. Horodecki \cite{phoro}. In Ref. \cite{upb} a method was discovered to construct 
entangled density matrices $\rho$ with positive semidefinite 
$\left({\rm id}_{A} \otimes T\right)(\rho)$ in various dimensions 
$\dim {\cal H}_A > 2$ and $\dim {\cal H}_B > 2$. The construction 
was based on the notion of an unextendible product basis.
Let us give the definition.

\begin{defi}
Let ${\cal H}$ be a finite dimensional Hilbert space
of the form ${\cal H}_A \otimes {\cal H}_B$. A partial product basis is a set ${\rm S}$ of mutually orthonormal pure
product states spanning a proper subspace ${\cal H}_{\rm S}$ of
${\cal H}$. An unextendible product basis is a partial product basis
whose complementary subspace ${\cal H}_{\rm S}^{\perp}$ contains no product
state.
\end{defi}

{\em Remark} This definition can be extended to product bases in 
${\cal H}=\bigotimes_{i=1}^m {\cal H}_i$ with arbitary $m$. Note we restrict 
ourselves to sets orthonormal sets S. 

With this notion we can construct the following density matrix:

\begin{theo}{\bf \cite{upb}}
Let ${\rm S}$ be a bipartite unextendible product basis 
$\{\ket{\alpha_i}\otimes \ket{\beta_i}\}_{i=1}^{|{\rm S}|}$ in 
${\cal H}={\cal H}_A \otimes {\cal H}_B$. We define a density matrix $\rho$ 
as
\beq
\rho=\frac{1}{\dim {\cal H}-|{\rm S}|}\left({\rm id}_{AB}-\sum_i \ket{\alpha_i} \bra{\alpha_i} \otimes 
 \ket{\beta_i} \bra{\beta_i}\right),
\label{boundrho}
\eeq
where ${\rm id}_{AB}$ is the identity on ${\cal H}$. The density matrix $\rho$ is entangled. Furthermore, the state $({\rm id}_A \otimes [{\cal S}_1 +T \circ {\cal S}_2])(\rho)$ for all completely positive maps ${\cal S}_1$ and ${\cal S}_2$, is positive semidefinite.
\label{upbbe}
\end{theo}

{\em Proof} The density matrix $\rho$ is 
proportional to the projector on the complementary subspace ${\cal H}_{\rm S}^{\perp}$. As S is unextendible  
${\cal H}_{\rm S}^{\perp}$ contains no product states. Therefore the density matrix
is entangled. It is not hard to see that $({\rm id}_A \otimes T)(\rho)$ is 
positive semidefinite. It has been proved in Ref. \cite{horodeckibound2} that when $({\rm id}_A \otimes T)(\rho)$ is positive semidefinite then $({\rm id}_A \otimes T \circ {\cal S}_2)(\rho)$ where ${\cal S}_2$ is any completely positive map, is also positive semidefinite. Therefore $({\rm id}_A \otimes [{\cal S}_1 +T \circ {\cal S}_2])(\rho)$ is also positive semidefinite. $\Box$

We are now ready to present our results relating these density matrices 
obtained from the construction in Theorem \ref{upbbe} to indecomposable positive linear maps. We will need the definition of a maximally entangled
pure state:

\begin{defi}
Let ${\cal H}={\cal H}_A \otimes {\cal H}_B$. Let $\ket{\psi}$ be a normalized
state in ${\cal H}$ and 
\beq
\rho_{A,\psi}={\rm Tr}_B \,\ket{\psi}\bra{\psi},
\eeq
where ${\rm Tr}_B$ indicates that the trace is taken with respect to 
Hilbert space ${\cal H}_B$ only. The state $\ket{\psi} \in {\cal H}$ is 
maximally entangled when
\beq
S(\rho_{A,\psi})=-{\rm Tr}\, \rho_{A,\psi} \log_2 \rho_{A,\psi}= \log_2 \min(\dim {\cal H}_A, \dim {\cal H}_B)
\label{vonneu}
\eeq  
\end{defi}
The function $S(\rho_{A,\psi})$ is the von Neumann entropy of the density matrix $\rho_{A,\psi}$.

{\em Remarks} For pure states $\ket{\psi}$ the von Neumann entropy of 
$\rho_{A,\psi}$ is always less than or equal to $d \equiv \log_2 \min(\dim {\cal H}_A, \dim {\cal H}_B)$. For maximally entangled states we will have
$\rho_{A,\psi}={\rm diag}(1/d,\ldots, 1/d,0,\ldots,0)$ so that the maximum von 
Neumann entropy, Eq. (\ref{vonneu}), is achieved. When $\dim {\cal H}_A=\dim {\cal H}_B$ one can always 
make an orthonormal basis for ${\cal H}$ with maximally entangled states \cite{fivel}.

The following lemma bounds the innerproduct between a maximally entangled state and any product state.

\begin{lem}
Let ${\cal H}={\cal H}_A \otimes {\cal H}_B$. Let $\ket{\Psi} \in {\cal H}$ 
be a maximally entangled state. Let $d=\min (\dim {\cal H}_A, \dim {\cal H}_B)$. For all (normalized) product states 
$\ket{\phi_A} \otimes \ket{\phi_B}$,
\beq
\left|\langle \Psi |\phi_A \rangle \otimes |\phi_B \rangle\right|^2 \leq \frac{1}{d}. 
\eeq
\label{distmax}
\end{lem}

{\em Proof} We write the maximally entangled state $\ket{\Psi}$ in the 
Schmidt polar form \cite{hughston} as 
\beq
\ket{\Psi}=\frac{1}{\sqrt{d}} \sum_{i=1}^d \ket{a_i} \otimes \ket{b_i}, 
\eeq
where $\langle a_i| a_j \rangle=\delta_{ij}$ and $\langle b_i| b_j \rangle=\delta_{ij}$. Thus we can write 
\beq
\left|\langle \Psi |\phi_A \rangle \otimes |\phi_B \rangle\right|^2=
\frac{1}{d}\left|\sum_{i=1}^d \langle \phi_A | a_i \rangle \langle \phi_B| b_i \rangle \right|^2 \leq \frac{1}{d},
\eeq
using the Schwarz inequality and $\sum_{i=1}^d |\langle \phi_A |a_i\rangle|^2 \leq 1$ and $\sum_{i=1}^d |\langle \phi_B |b_i\rangle|^2 \leq 1$. $\Box$

We will also need the following lemma:

\begin{lem}
Let ${\rm S}$ be an unextendible product basis $\{\ket{\alpha_i} \otimes \ket{\beta_i}\}_{i=1}^{|{\rm S}|}$ in ${\cal H}={\cal H}_A \otimes {\cal H}_B$. Let 
\beq
f(\ket{\phi_A},\ket{\phi_B})=\sum_{i=1}^{|{\rm S}|} |\langle \phi_A | \alpha_i \rangle|^2 
|\langle \phi_B |\beta_i \rangle |^2.
\label{epsfid}
\end{equation}
The minimum of $f$ over all pure states $\ket{\phi_A} \in {\cal H}_A$ and 
$\ket{\phi_B} \in {\cal H}_B$ exists and is strictly larger than 0.
\label{eexact}
\end{lem}

{\em Proof} The set of all pure product states $\ket{\phi_A} \otimes \ket{\phi_B}$ on ${\cal H}$ is a compact set. The function $f$ is a continuous function on this set. Therefore, if 
there exists a set of states $\ket{\phi_A} \otimes \ket{\phi_B}$ for which $f$ is arbitrary 
small then there would also exist a pair $\ket{\phi'_A} \otimes \ket{\phi'_B}$ for which $f=0$. This contradicts the fact that S is an unextendible product basis. $\Box$

The following two theorems contain the main result of this paper. 

\begin{theo}
Let ${\rm S}$ be an unextendible product basis $\{\ket{\alpha_i} \otimes \ket{\beta_i}\}_{i=1}^{|{\rm S}|}$ in ${\cal H}={\cal H}_A \otimes {\cal H}_B$. Let $\rho$ 
be the density matrix 
\beq
\rho=\frac{1}{\dim {\cal H}-|{\rm S}|}\left({\rm id}_{AB}-\sum_{i=1}^{|{\rm S}|} \ket{\alpha_i} \bra{\alpha_i} \otimes 
 \ket{\beta_i} \bra{\beta_i}\right),
\label{densmat}
\eeq
Let $d=\min(\dim {\cal H}_A, \dim {\cal H}_B)$. Let ${\rm H}$ be a Hermitian 
operator given by
\beq
{\rm H}=\sum_{i=1}^{|{\rm S}|} \ket{\alpha_i}\bra{\alpha_i} \otimes \ket{\beta_i}\bra{\beta_i}- d \eps \ket{\Psi}\bra{\Psi}, 
\label{defh}
\eeq
where $\ket{\Psi}$ is a maximally entangled state such that
\beq
\langle \Psi |\,\rho\,|\Psi \rangle > 0,
\label{inprod}
\end{equation}
and 
\beq
\eps=\min_{\ket{\phi_A} \otimes \ket{\phi_B}} \sum_{i=1}^{|{\rm S}|} |\langle \phi_A | \alpha_i \rangle|^2 
|\langle \phi_B |\beta_i \rangle |^2,
\label{epseq}
\end{equation}
where the minimum is taken over all pure states $\ket{\phi_A} \in {\cal H}_A$ and $\ket{\phi_B} \in {\cal H}_B$. For any unextendible product basis ${\rm S}$ it is possible to find a maximally entangled state $\ket{\Psi}$ such that Eq.(\ref{inprod}) holds. 
${\rm H}$ has the following properties: 
\beq
{\rm Tr}\, {\rm H}\, \rho < 0,
\label{negent}
\end{equation}
and for all product states $\ket{\phi_A} \otimes \ket{\phi_B} \in {\cal H}$, 
\beq
{\rm Tr}\, {\rm H} \ket{\phi_A} \bra{\phi_A} \otimes \ket{\phi_B} \bra{\phi_B} \geq 0. 
\label{posmap}
\end{equation}
\label{witness}
\end{theo}

{\em Proof} Eq. (\ref{posmap}) follows from the definition of $\eps$, Eq. (\ref{epseq}), and 
Lemma \ref{distmax}. Consider Eq. (\ref{negent}). As the density matrix 
$\rho$ is proportional to the projector on ${\cal H}_{\rm S}^{\perp}$, one has
\beq
{\rm Tr}\, {\rm H}\, \rho =-d \eps \,\bra{\Psi}\, \rho\, \ket{\Psi}, 
\end{equation}
which is strictly smaller than zero by Lemma \ref{eexact} and the choice of the maximally entangled
state, Eq. (\ref{inprod}). When $\dim {\cal H}_A=\dim {\cal H}_B$ there 
exist bases of maximally entangled states and thus there will be a basis vector
$\ket{\Psi}$ for which $\bra{\Psi}\, \rho\, \ket{\Psi}$ is nonzero.
In case, say, $\dim {\cal H}_A > \dim {\cal H}_B$, the maximally entangled 
states form bases of subspaces ${\cal H}'={\cal H}_A' \otimes {\cal H}_B$ 
with ${\cal H}_A' \subset {\cal H}_A$ and $\dim {\cal H}_A'=\dim {\cal H}_B$.
This completes the proof. $\Box$

\begin{theo}
Let ${\rm S}$ be an unextendible product basis $\{\ket{\alpha_i} \otimes \ket{\beta_i}\}_{i=1}^{|{\rm S}|}$ in ${\cal H}={\cal H}_A \otimes {\cal H}_B$.
Let ${\rm H}$ be defined as in Theorem \ref{witness}, Eq. (\ref{defh}). Choose an orthonormal basis $\{\ket{i}\}_{i=1}^{\dim {\cal H}_A}$ for ${\cal H}_A$. Let ${\cal S} \colon\; B({\cal H}_A) \rightarrow B({\cal H}_B)$ be a linear map defined by 
\beq
{\cal S}\,(\ket{i}\bra{j})=\langle i|\,{\rm H}\,|j\rangle.
\label{equival}
\eeq 
Then ${\cal S}$ is positive but not completely positive. ${\cal S}$ is indecomposable.
\label{indecom}
\end{theo}

%{\em Remark} The map ${\cal S}$ is uniquely determined by ${\rm H}$ modulo  
%basis transformations in ${\cal H}_A$.
 
{\it Proof} The relation between ${\cal S}$ and ${\rm H}$, Eq. (\ref{equival}), 
follows from the isomorphism between Hermitian operators on 
${\cal H}_A \otimes {\cal H}_B$ with the property of Eq. (\ref{posmap}) and 
linear positive maps, see \cite{nec_horo,jami}. In particular, iff 
a Hermitian ${\rm H}$ operator on ${\cal H}_A \otimes {\cal H}_B$ has the 
property of Eq. (\ref{posmap}) then the linear map ${\cal R}\colon\;
B({\cal H}_A) \rightarrow B({\cal H}_B)$ defined by 
\beq
{\rm H}=\sum_{i,j} (\ket{i}\bra{j})^{*} \otimes {\cal R}(\ket{i}\bra{j}),
\eeq
is positive for any choice of the orthonormal basis $\{\ket{i}\}_{i=1}^{\dim {\cal H}_A}$.
The map ${\cal S}={\cal R} \circ T$ where $T$ is matrix transposition 
in the basis $\{\ket{i}\}_{i=1}^{\dim {\cal H}_A}$ of Eq. (\ref{equival}) 
is then positive as well.

We will show how the density matrix $\rho$ derived from the unextendible
product basis, Eq. (\ref{densmat}) shows that ${\cal S}$ is not completely 
positive. At the same time we prove that the assumption that ${\cal S}$ is 
decomposable leads to a contradiction. We can rewrite Eq. (\ref{equival}) as 
\beq
{\rm H}=({\rm id}_A \otimes {\cal S})(\ket{\Psi^+}\bra{\Psi^+})
\label{inbet}
\eeq
where $\ket{\Psi^+}$ is equal to the (unnormalized) maximally entangled state 
$\sum_{i=1}^{\dim {\cal H}_A} \ket{i} \otimes \ket{i}$. Let ${\cal S}^*\colon\; B({\cal H}_B) \rightarrow B({\cal H}_A)$ be the Hermitian conjugate of 
${\cal S}$. We use the definition of ${\cal S}^*$ 
\beq
{\rm Tr}\, {\cal S}^*(A^*)\, B={\rm Tr} A^{*} \,{\cal S}(B),
\end{equation}
and Eq. (\ref{inbet}) to derive that Eq. (\ref{negent}) can be rewritten as 
\beq
{\rm Tr}\, {\rm H} \,\rho =\bra{\Psi^{+}} \left({\rm id}_A \otimes {\cal S}^*\right)(\rho) \ket{\Psi^{+}} < 0, 
\label{rewrit}
\eeq
Thus ${\cal S}^*$ cannot be completely positive and therefore ${\cal S}$ 
itself is not completely positive. If ${\cal S}$ were decomposable, then 
${\cal S}^*$ would be of the form ${\cal S}_1+T \circ {\cal S}_2$ where 
${\cal S}_1$ and ${\cal S}_2$ are completely positive maps. The density 
matrix $\rho$ is  positive semidefinite under any linear map of the 
form ${\cal S}_1+T \circ {\cal S}_2$ by Theorem \ref{upbbe}. This is in contradiction with Eq. (\ref{rewrit}) and therefore ${\cal S}$ cannot be decomposable. $\Box$

We will now show how one can determine a lower bound on the value of $\epsilon$
, Eq. (\ref{epseq}). Note that when we determine a lower bound $\eps \geq 
\eps_{max}$, then all operators ${\rm H}$, as in Eq. (\ref{defh}) of the form 
\beq
{\rm H}=\sum_{i=1}^{|{\rm S}|} \ket{\alpha_i}\bra{\alpha_i} \otimes \ket{\beta_i}\bra{\beta_i}- d \mu \ket{\Psi}\bra{\Psi}, 
\eeq
where $\mu \in (0,\eps_{max}]$ correspond to positive maps. 

Let $\{\ket{\alpha_i} \otimes \ket{\beta_i}\}_{i=1}^{|{\rm S}|}$ 
be an unextendible product basis in ${\cal H}_A \otimes {\cal H}_B$ with 
$d_A=\dim {\cal H}_A$ and $d_B=\dim {\cal H}_B$. Let ${\rm S}_A=\{\ket{\alpha_i}\}_{i=1}^{|{\rm S}|}$ and ${\rm S}_B=\{\ket{\beta_i}\}_{i=1}^{|{\rm S}|}$. 
We pick a vector $\ket{\phi_A}$ and order the innerproducts 
$|\langle \alpha_i| \phi_A \rangle|^2$ in an increasing sequence; let us 
call them $x_1 \leq x_2 \leq \ldots \leq x_{|{\rm S}|}$. Then we select
vectors $\ket{\alpha_i}$ corresponding to the smallest innerproducts in this sequence up to
the point where the set of selected vectors $\ket{\alpha_i}$ spans the 
 full $d_A$-dimensional Hilbert space ${\cal H}_A$. Let us call this set 
${\rm S}_A^P\in {\rm S}_A$. If we would take away anyone state from ${\rm S}_A^P$, the remaining vectors would no longer span ${\cal H}_A$. As the vectors in the set ${\rm S}_A^P$ span ${\cal H}_A$, it must be that $x_{|{\rm S}_A^P|} > 0$. Let us 
label this corresponding vector as $\ket{\alpha_{i_{max}}}$, i.e. 
$x_{|{\rm S}_A^P|}=|\langle \alpha_{i_{max}}| \phi_A \rangle|^2$. Now we construct a subset of $S_B$ in the following way; we define ${\rm S}_B^P= \{\ket{\beta_i}\,|\, \ket{\alpha_i} \not \in {\rm S}_A^P\} \cup \{\ket{\beta_{i_{max}}}\}$. 
We note that the vectors in the set ${\rm S}_B^P$ span the Hilbert space ${\cal H}_B$; if not, then there would exist a vector $\ket{\phi_B}$ which is 
orthogonal to all vectors in ${\rm S}_B^P$ {\em and} a vector $\ket{\phi_A}$ 
which is orthogonal to all vectors in ${\rm S}_A^P \setminus \{\ket{\alpha_{i_{max}}}\}$, 
wich would in turn imply that $\eps=0$, in other words the set S would be 
extendible. Let us pick a vector $\ket{\phi_B}$ and denote 
the innerproducts $|\langle \beta_i| \phi_B \rangle|^2$ with $\ket{\beta_i} \in {\rm S}_B^P$ as $y_1 \leq y_2 \leq \ldots \leq y_{|{\rm S}_B^P|}$. 
As the vectors in ${\rm S}_B^P$ span ${\cal H}_B$, we know that 
$y_{|{\rm S}_B^P|} > 0$ for any state $\ket{\phi_B}$. This implies that for 
a particular $\ket{\phi_A}$ and $\ket{\phi_B}$ we can bound 
\beq
\sum_i |\langle \alpha_i|\phi_A \rangle|^2 |\langle \beta_i|\phi_B \rangle|^2
\geq x_{|{\rm S}_A^P|} \; y_{|{\rm S}_B^P|},
\label{lowerb}
\eeq
the product of the two largest innerproducts of the vectors $\ket{\phi_A}$ and 
$\ket{\phi_B}$ with the vectors from ${\rm S}_A^P$ and ${\rm S}_B^P$ respectively. 
Therefore $\eps$ itself, Eq. (\ref{epseq}), can be bounded as 
\beq
\eps \geq \min_{\ket{\phi_A}->{\rm S}_A^P,\ket{\phi_B}->{\rm S}_B^P}  x_{|{\rm S}_A^P|} y_{|{\rm S}_B^P|},
\label{epscompl}
\eeq
where $x_{|{\rm S}_A^P|}$ denotes the largest innerproduct between 
$\ket{\phi_A}$ and a state in the set $S_A^P$. We minimize over ${\ket{\phi_A}->{\rm S}_A^P}$ and $\ket{\phi_B}->{\rm S}_B^P$ where the arrow denotes that 
a state $\ket{\phi_A}$ gives rise to a set $S_A^P$ as in the construction 
given above. A set $S_A^P$ (and similarly $S_B^P$) might not be uniquely defined given the vector $\ket{\phi_A}$, for example when several 
innerproducts of $\ket{\phi_A}$ with states $\ket{\alpha_i}$ are identical. 
Since the lowerbound given in Eq. (\ref{lowerb}) works for all sets $S_A^P$ and $S_B^P$ which are constructed with the method given above, we could do 
an extra maximization over $S_A^P$ and $S_B^P$, given the states $\ket{\phi_A}$ and 
$\ket{\phi_B}$, but for the sake of clarity this maximization is omitted in 
Eq. (\ref{epscompl}).

We have the following proposition that can be used to bound 
$x_{|{\rm S}_A^P|}$ and $y_{|{\rm S}_B^P|}$ given the sets ${\rm S}_A^P$ and ${\rm S}_B^P$:

\begin{propo} 
Let $\{\ket{\psi_i}\}_{i=1}^{n}$ be a set of $n$ vectors in ${\cal H}$ such that 
$\{{\ket{\psi_i}}\}_{i=1}^{n}$ span the Hilbert space ${\cal H}$.
Then for any vector $\ket{\phi}$ we have 
\beq
n \max_i |\bra{\phi} \psi_i\rangle|^2 \geq \sum_i |\bra{\phi} \psi_i\rangle|^2 \geq \lambda_{min},  
\eeq
where $\lambda_{min}$ is the smallest eigenvalue of the Hermitian matrix 
$P=\sum_i \ket{\psi_i} \bra{\psi_i}$. 
\label{evalmin}
\end{propo}

Summarizing, we get the following 
\beq
\eps \geq \min_{{\rm S}_B^P,{\rm S}_B^P} \frac{\lambda_{min,{\rm S}_A^P}}{|{\rm S}_A^P|} \frac{\lambda_{min,{\rm S}_B^P}}{|{\rm S}_B^P|} \equiv \eps_{max}.
\label{epslower}
\eeq
In order to carry out this calculation, we first find all minimal `full rank' subsets ${\rm S}_A^P$ of ${\rm S}_A$. Then for each of these 
sets ${\rm S}_A^P$ we compute the smallest eigenvalue of $\sum_{i \in {\rm S}_A^P} \ket{\alpha_i} \bra{\alpha_i}$. Also for each set ${\rm S}_A^P$, we construct complementary sets 
 ${\rm S}_B^P$ which contain all the vectors $\ket{\beta_i}$ such that 
$\ket{\alpha_i} \not \in {\rm S}_A^P$ and a single state $\ket{\beta_i}$ such $\ket{\alpha_i} \in {\rm S}_A^P$. For each set ${\rm S}_A^P$ there will be 
$|{\rm S}_A^P|$ of such sets ${\rm S}_B^P$. Then for each ${\rm S}_B^P$ we compute the smallest eigenvalue of $\sum_{i \in {\rm S}_B^P} \ket{\beta_i} \bra{\beta_i}$. Then we can take the minimum over all these values to obtain a bound on $\eps$. Note that this is now a minimization over a discrete number of values.
If the set S has few symmetries and is defined in a high dimensional space, 
the procedure will be cumbersome. In small dimensions or for unextendible 
product bases which do have many symmetries, the calculation will be 
much simpler. In the next section we present two examples of positive maps based on the construction in Theorem \ref{indecom} and for one of them we will 
explicitly compute a lower bound on $\epsilon$.

\section{Examples and Discussion}
\label{exampl}

As we have shown the structure of unextendible product bases carries over to 
 indecomposable positive linear maps. In this section we will list some of the 
results that have been obtained about unextendible product bases. We will 
take two examples of unextendible product bases and demonstrate the 
construction of Theorem \ref{witness} and Theorem \ref{indecom}.

\begin{enumerate}
\item In Ref. \cite{upb} it was shown that there exist no unextendible 
product bases in ${\cal H}_2 \otimes {\cal H}_n$ for any $n \geq 2$.
\item In Ref. \cite{upbbe2} it was shown how to parametrize {\em all}  
possible unextendible product bases in ${\cal H}_3 \otimes {\cal H}_3$ as a 
six-parameter family.
\item In Ref. \cite{upbbe2} a family of unextendible product bases, 
based on quadratic residues, in ${\cal H}_n \otimes {\cal H}_n$ where $n$ is any odd number and $2n-1$ is a prime of the form $4m+1$ has been found. 
\item In Ref. \cite{upbbe2} a family of unextendible product bases 
${\cal H}_n \otimes {\cal H}_m$ ($m > 2$, $n > 2$) for
arbitary $m \neq n$ as well as even $n=m$ has been conjectured. The conjecture 
was proved in ${\cal H}_3 \otimes {\cal H}_n$ and ${\cal H}_4 \otimes {\cal H}_4$ (see also \cite{upbsdivterhal}).
\item In Ref. \cite{upbbe2} it was shown that when ${\rm S}_1$ and ${\rm S}_2$ are 
unextendible product bases on ${\cal H}_A^1 \otimes {\cal H}_B^1$ and
${\cal H}_A^2 \otimes {\cal H}_B^2$ respectively, then the tensorproduct of 
the two sets, ${\rm S}_1 \otimes {\rm S}_2$, is again an unextendible 
product bases on $({\cal H}_A^1 \otimes {\cal H}_A^2) \otimes ({\cal H}_B^1 \otimes 
 {\cal H}_B^2)$.
\end{enumerate}

{\bf Example 1}: One of the first examples of an unextendible product basis
in ${\cal H}_3 \otimes {\cal H}_3$ was the following set of states \cite{upb}.
Consider five vectors in real three-dimensional space forming the apex of 
a regular pentagonal pyramid, the height $h$ of the pyramid being chosen 
such that nonadjacent apex vectors are orthogonal. The vectors are  
\beq
\ket{v_i}=N (\cos{{ 2 \pi i \over 5}}, \sin{{2 \pi i}\over 5},h), \;\; 
i=0,\ldots,4,
\label{defP}
\end{equation}
with $h={1 \over 2} \sqrt{1+\sqrt{5}}$ and $N=2/\sqrt{5+\sqrt{5}}$. 
Then the following five states in ${\cal H}_3 \otimes {\cal H}_3$ form 
an unextendible product basis:
\beq
\ket{p_i} = \ket{v_i} \otimes  \ket{v_{2i \bmod 5}}, \;\; i=0,\ldots,4.
\label{defpent}
\end{equation}
Let $\rho$ be the entangled state derived from this unextendible product 
basis as in Eq. (\ref{boundrho}) Theorem \ref{upbbe}. We choose a 
maximally entangled state $\ket{\Psi}$, here named $\ket{\Psi^+}$,
\beq
\ket{\Psi^{+}}=\frac{1}{\sqrt{3}}(\ket{11}+\ket{22}+\ket{33}).
\label{defpsiplus}
\end{equation}
One can easily compute that 
\beq
\bra{\Psi^{+}}\,\rho\,\ket{\Psi^{+}}=\frac{1}{4}\left(1-\frac{7+\sqrt{5}}{3(3+\sqrt{5})}\right) > 0.
\end{equation}
Let us now compute a lower bound on $\eps$, as in Eq. (\ref{epslower}). Due to the high symmetry of this set
of states, we will only need the compute the minimum eigenvalue of the 
Hermitian matrix $P_1=\ket{v_0}\bra{v_0}+\ket{v_1}\bra{v_1}+\ket{v_2}\bra{v_2}$ and 
 $P_2=\ket{v_0}\bra{v_0}+\ket{v_1}\bra{v_1}+\ket{v_3}\bra{v_3}$; all other 
subsets of three vectors, either on Bob's or Alice's side, correspond to 
matrices with the same eigenvalues as $P_1$ or $P_2$.
Easy computation shows that $P_1$ has the smallest eigenvalue which 
is equal to 
\beq
\lambda_{min}=\frac{2+\sqrt{2}-\sqrt{10}}{2}.
\eeq
Then as the states on Bob's side are identical, we get 
\beq
\eps \geq \frac{\lambda_{min}^2}{9}=\frac{4+\sqrt{2}-\sqrt{5}-\sqrt{10}}{9}.
\eeq
The map ${\cal S}$ as defined in Eq. (\ref{equival}) Theorem \ref{indecom}, 
follows directly:
\beq
{\cal S}(\ket{i}\bra{j})=\sum_{k=0}^4 \langle i|v_{k}\rangle \langle v_{k}|j \rangle \ket{v_{2k \bmod 5}} \bra{v_{2k \bmod 5}}-3\mu \ket{i} \bra{j}.
\eeq
where 
\beq
\mu \in \left(0,\frac{4+\sqrt{2}-\sqrt{5}-\sqrt{10}}{9}\right].
\eeq

A positive linear map ${\cal S} \colon \;B({\cal H}_n) \rightarrow B({\cal H}_m)$ is unital if ${\cal S}({\rm id}_n)={\rm id}_m$. We will demonstrate that ${\cal S}$ is not unital.
One can write
\beq
{\cal S}({\rm id}_A)={\rm Tr}_A \,{\rm H}=\sum_{k=0}^{4} 
\ket{v_{2k \bmod 5}} \bra{v_{2k \bmod 5}}-3 \mu\; {\rm Tr}_A \ket{\Psi^+} \bra{\Psi^+},
\label{unit}
\eeq
which in turn is equal to 
\beq
{\cal S}({\rm id}_A)={\rm diag}\left(\frac{10}{5+\sqrt{5}},\frac{10}{5+\sqrt{5}},\sqrt{5}\right)-\mu \,{\rm id}_B.
\eeq

The next example is based on a more general unextendible product bases
that was presented in Ref. \cite{upbbe2}.

{\bf Example 2}:
The states of S in ${\cal H}_3 \otimes {\cal H}_n$ are
\bea
\ket{F_k^0}&=&\frac{1}{\sqrt{n-2}}|0\rangle \otimes (|1\rangle+\sum_{l=3}^{n-1}\omega^{k(l-2)}|l\rangle),\ \ \
1\leq k\leq n-3,\\
\ket{F_k^1}&=&\frac{1}{\sqrt{n-2}} |1\rangle \otimes (|2\rangle+\sum_{l=3}^{n-1}\omega^{k(l-2)}|l\rangle),\ \ \
1\leq k\leq n-3,\\
\ket{F_k^2}&=&\frac{1}{\sqrt{n-2}}|2\rangle \otimes (|0\rangle+\sum_{l=3}^{n-1}\omega^{k(l-2)}|l\rangle),\ \ \
1\leq k\leq n-3,\\
\ket{\psi_3}&=&\frac{1}{\sqrt{2}}(\ket{0}-\ket{1}) \otimes |0\rangle,\\
\ket{\psi_4}&=&\frac{1}{\sqrt{2}}(\ket{1}-\ket{2})\otimes |1\rangle,\\
\ket{\psi_5}&=&\frac{1}{\sqrt{2}}(\ket{2}-\ket{0})\otimes |2\rangle,\\
\ket{\psi_6}&=&\frac{1}{\sqrt{3n}}\sum_{i=0}^2\sum_{j=0}^{n-1}|i\rangle\otimes |j\rangle,
\eea
and we have $\omega=\exp(2\pi i/(n-2))$. Here the states 
$\{\ket{k}\}_{k=0}^{n-1}$ form an orthonormal basis. In total there are 
$3n-5$ states in the basis.
We choose a maximally entangled state, again we take $\ket{\Psi^+}$, Eq. (\ref{defpsiplus}). One can show that 
\beq
\langle \Psi^+ |\,\rho\,| \Psi^+\rangle=\frac{1}{5}\left(\frac{1}{2}-\frac{1}{3n}\right) > 0.
\eeq 

The map ${\cal S}\colon\, B({\cal H}_3) \rightarrow B({\cal H}_n)$ is given 
as 
\beq
{\cal S}(\ket{i}\bra{j})=\sum_{k=1}^{n-3} \sum_{p=0}^2 \langle i | F_k^p \rangle \langle F_k^p | j \rangle+\sum_{p=3}^6 \langle i | \psi_p \rangle \langle \psi_p | j \rangle - \eps \,\ket{i} \bra{j}.
\eeq

The following questions concerning the positive maps that 
were introduced in this paper are left open. 
\begin{enumerate}
\item Is ${\cal S}$ always non unital? We conjecture it is. As we showed, see
Eq. (\ref{unit}), the answer to this question depends on whether 
\beq
\sum_{i=1}^{|{\rm S}|} \ket{\beta_i} \bra{\beta_i} \propto {\rm id}_B, 
\eeq
where the set of states $\{\ket{\beta_i}\}_{i=1}^{|{\rm S}|}$ are one side of 
the unextendible product basis. The states $\ket{\beta_i}$ will span ${\cal H}_B$
but they will not be all orthogonal, nor all nonorthogonal.
\item It was shown in Theorem \ref{indecom} that the new indecomposable positive linear maps 
${\cal S}\colon\, B({\cal H}_m) \rightarrow B({\cal H}_n)$ are not $m$-positive, as they are not completely positive. Are these maps ${\cal S}$ $k$-positive with $1 < k < m$?  
The answer to this question will rely on a better understanding of the 
structure of unextendible product bases.
\item In \cite{upb} a single example was given of a entangled density matrix on
${\cal H}_3 \otimes {\cal H}_4$ which was positive semidefinite under 
${\rm id_3} \otimes T$. The density matrix was based not on an unextendible 
product basis, but a `strongly uncompletable' product basis S. It could be shown that the 
Hilbert space ${\cal H}_{\rm S}^{\perp}$ had a product state {\em deficit}, i.e.
the number of product states in ${\cal H}_{\rm S}^{\perp}$ was less than 
$\dim {\cal H}_{\rm S}^{\perp}$. It is open question how to generalize this 
example and whether these kinds of density matrices will give rise to more general
 family of indecomposable positive linear maps. 
\end{enumerate}

{\em The author would like to thank David DiVincenzo for fruitful discussions,
 Alan Hoffman for suggestions to improve the text and Peter Shor for 
sharing his insights in how to derive a lower bound on $\epsilon$.}


\begin{references}
\bibitem{upbsdivterhal}The full conjecture (arbitrary $n$ and $m$) has recently been proved by Terhal and 
DiVincenzo and will be presented in a forthcoming paper.
\end{references}
\end{document}